

\def\l{\lambda}

\def\r{\rho}
\def\s{\sigma}

 \def\noss{\noalign{\smallskip}}
\def\noms{\noalign{\medskip}}

\def\dal{\sqcup\kern-.29cm{\sqcap}}

\def\nofirstpagenoten{\nopagenumbers\footline={\ifnum\pageno>1\tenrm
\hss\folio\hss\fi}}
\def\nofirstpagenotwelve{\nopagenumbers\footline={\ifnum\pageno>1\twelverm
\hss\folio\hss\fi}}
\def\nofirstpagenotwelver{\nopagenumbers\footline={\ifnum\pageno>0\twelverm
\hss\folio\hss\fi}}

\def\ft#1#2{{\textstyle{{#1}\over{#2}}}}
\def\frac#1#2{{{#1}\over{#2}}}
\def\1#1{\frac1{#1}} \def\2#1{\frac2{#1}} \def\3#1{\frac3{#1}}
\def\noss{\noalign{\smallskip}}
\def\sb#1{\lower.4ex\hbox{${}_{#1}$}}

\def\pa{\partial}

\def\'{\mkern 1mu}

\def\crampest{\medmuskip = 1mu plus 1mu minus 1mu}
\def\uncramp{\medmuskip = 4mu plus 2mu minus 4mu}

\def\IR{{\hbox{{\rm I}\kern-.2em\hbox{\rm R}}}}
\def\IB{{\hbox{{\rm I}\kern-.2em\hbox{\rm B}}}}
\def\IN{{\hbox{{\rm I}\kern-.2em\hbox{\rm N}}}}
\def\IC{{\ \hbox{{\rm I}\kern-.6em\hbox{\bf C}}}}

\def\IZ{{\hbox{{\rm Z}\kern-.4em\hbox{\rm Z}}}}
\def\to{\rightarrow}

\def\underarrow#1{\vbox{\ialign{##\crcr$\hfil\displaystyle
{#1}\hfil$\crcr\noalign{\kern1pt
\nointerlineskip}$\longrightarrow$\crcr}}}
%

\nofirstpagenotwelve
\pageno=0
\input phyzzx
\tolerance=5000
\overfullrule=0pt

\twelvepoint
\pubnum{CERN-TH.6633/92}
\date{\ }
\bigskip
\vfil
\titlepage
\title{QUANTUM DILATON GRAVITY IN THE LIGHT--CONE GAUGE}
\vglue-.25in
\medskip
\author{X. Shen\foot{Work supported partially by a World Laboratory
scholarship.\hfil\break
\medskip
{\noindent{\twelvepoint CERN-TH.6633/92\hfill\break
September 1992\hfill}}
}}
\medskip
\address{Theory Division, CERN
\break CH-1211, Geneva 23, Switzerland}
\bigskip
\abstract{Recently, models of two-dimensional dilaton gravity have been
shown to admit classical black-hole solutions that exhibit Hawking
radiation at the semi-classical level. These classical and semi-classical
analyses have been performed in conformal gauge. We show in this paper
that a similar analysis in the light--cone gauge leads to the same
results. Moreover, quantization of matter fields in light--cone gauge
can be   naturally extended to include quantizing the metric field
{\it \`a la} KPZ.  We argue that this may provide a new framework to
address many issues associated to black-hole physics.}

\endpage

\REF\gsw{M.~ Green, J.~H.~ Schwarz and E.~ Witten, Superstring theory
(Cambridge University Press, Cambridge, 1987).}
\REF\haw{S.~W.~ Hawking, Commun. Math. Phys. {\bf 43} (1975) 199.}
\REF\hawk{S.~W.~ Hawking, Phys. Rev. {\bf D14} (1976) 2460;\  for a recent
review, see R. Wald, lectures given at 1991 Erice.}
\REF\witten{E.~ Witten, Phys. Rev. {\bf D44} (1991) 314.}
\REF\cghs{C.~G.~ Callan Jr., S.~B.~ Giddings, J.~A.~ Harvey and
A.~ Strominger, Phys. Rev. {\bf D45} (1992) R1005.}
\REF\rst{J.~G.~ Russo, L.~ Susskind and L.~ Thorlacius, Stanford
University preprint SU-ITP-92--4.}
\REF\st{L.~ Susskind and L.~ Thorlacius, Stanford University
preprint SU-ITP-92--12.}
\REF\bddo{T.~ Banks, A.~ Dabholkar, M.~R.~ Douglas and M.~ O'Loughlin,
Rutgers preprint RU-91--54 (January, 1992).}
\REF\hawkk{S.~W.~ Hawking, Caltech preprint CALT-68-1774,
hepth@xxx/923052.}
\REF\bghs{B.~ Birnir, S.~B.~ Giddings, J.~A.~ Harvey and A.~ Strominger,
UCSB/Chicago preprint UCSBTH-92-08=EFI-92-16, hepth@xxx/9203042.}
\REF\stro{A.~ Strominger, Santa Barbara ITP preprint
UCSBTH-92--18, hep@xxx/9205028.}
\REF\sb{S.~B.~ Giddings and A.~ Strominger, Santa Barbara preprint
UCSBTH-92-28, hepth@xxx/9207034.}
\REF\alwiss{S.~P.~ de Alwis, CERN-TH.COLO-HEP-288,
hepth@xxx/9207095.}
\REF\hs{S.~W.~ Hawking and J.~M.~ Stewart, DAMTP
preprint, hepth@xxx/9207105, July 1992.}
\REF\cb{C.~ Callan and A.~ Bilal, Princeton University preprint PUPT-1320,
May 1992, hepth@xxx/9205089.}
\REF\alwis{S.~P.~ de Alwis, University of Colorado
preprint COLO-HEP-280, hepth@xxx/9205069, May 1992.}
\REF\rstt{J.G. Russo, L. Susskind and L. Thorlacius, Stanford
University preprint SU-ITP-92--17, June 1992.}

\REF\sltwo{A.~M.~ Polyakov, Mod. Phys. Lett. {\bf A2} (1987) 893.}
\REF\kpz{V.~G.~ Knizhnik, A~.M.~ Polyakov and A.~B.~ Zamolodchikov, Mod.
Phys. Lett. {\bf A3} (1988) 819.}
\REF\tanii{Y.~ Tanii, preprint STUPP-92-130, August 1992}
\REF\ddk{F.~ David, Mod. Phys. Lett. {\bf A3} (1988) 1651; \hfil\break
J.~ Distler and H.~ Kawai, Nucl. Phys. {\bf B321} (1989) 509.}
\REF\wgrav{M.~ Bershadsky and H.~ Ooguri, Commun. Math. Phys. {\bf 126}
(1989)49; \hfil\break K.~ Schoutens, A.~ Sevrin and P.~ van~ Nieuwenhuizen,
contribution to Trieste Summer School on High Energy Physics and
Cosmology, Trieste, Italy, Jun 17-Aug 9, 1991; \hfil\break K.~ Schoutens,
A.~ Sevrin  and P.~ van~ Nieuwenhuizen, Nucl. Phys. {\bf B364}
(1991);\hfil\break   M.~ Grisaru and P.~ van~ Nieuwenhuizen,
CERN-TH.6388/92; \hfil\break  X.~ Shen, preprint CERN-TH.6404/92.}
\REF\poly{A.~M.~ Polyakov, in Proceedings of the Les Houches 1988 meeting
on {\it Fields, Strings and Critical Phenomena} (North-Holland,
Amsterdam, 1989).}
\REF\pk{K.~A.~ Meisner and J.~ Pavelchik, Mod. Phys. Lett. {\bf A5} (1990)
763.}
\REF\emn{J.~ Ellis, D.~V.~ Nonapolous and N.~E.~ Mavromatos, Phys. Lett.
{\bf B267} (1991) 465; {\bf B272} (1991) 261; CERN-TH.6595/92, and the
references therein; \hfil\break  F.~ Yu and Y.~S.~ Wu, Phys. Rev. Lett.
{\bf 68} (1992) 2996; \hfil\break S.~ Chaudhuri and J.~D.~ Lykken,
preprint FERMI-PUB-92/169--T.}
\REF\nooda{S.~ Nojiri and I.~ Oda, preprint NDA-FP-4/92, OCHA-PP-26,
June 1992.}

\thispage1
\bigskip
\noindent{\bf 1.\  Introduction}

The reconciliation between the principles of quantum mechanics and
general relativity has been a long-standing challenge in theoretical
physics. Conventional quantization of Einstein's general relativity leads
to unrenormalizability of the theory and thus to the breakdown of any
predictive power. An indirect way to quantize gravity came
surprisingly from an unsuccessful attempt to  describe strong interactions
in terms of string dynamics. It is generally believed  that a consistent
quantum theory of gravity, as well as all other gauge interactions,
emerges in the framework of string theory [\gsw].

Prior to the advent of modern string theory, in the seminal work of Hawking
[\haw], it has been shown that a semi-classical treatment of
four-dimensional black holes---one of the focal
points of quantum gravity---leads to Hawking radiation. However this leaves
a series of issues that beg for further elucidation, such as
quantum coherence, back reaction on the  metric from the radiation,
evolution of singularity, and the endpoint of black-hole evaporation [\hawk].
Naturally one may ask if the string picture of quantum  gravity can offer
some insights in these issues. In [\witten], Witten  obtained a
two-dimensional black-hole solution as an exact solution of string theory,
which triggered a great deal of interest in studying black holes from the
viewpoints of string theory. It was conjectured [\witten] that the endpoint
of Hawking radiation for this two-dimensional black hole should be
described by the matrix model formulation of two-dimensional string theory
($c=1$).

Following Witten's discovery of two-dimensional stringy black hole, an
interesting two-dimensional model was proposed by Callan, Giddings,
Harvey and Strominger (CGHS) [\cghs] to describe the dynamics of Hawking
radiation. This model admits a classical solution of the black hole, with
a singularity shielded by horizon, and at semi-classical level exhibits
Hawking radiation. It is therefore believed that it serves as a promising
toy model to discuss issues of back reaction, quantum coherence,
singularity, and the fate of black hole. There have been a lot of
investigations of this model [\rst--\hs], as well as some  well-motivated
variants [\cb--\rstt]. It has been shown in these models that Hawking
radiation at semi-classical level leads to either a naked singularity
[\rst--\hawkk] or thunderbolt [\hs].

The analysis carried out in the original CGHS model, and later in its
variants, has been performed in conformal gauge. There the
classical equations of motion for dilaton, conformal field of gravity
and matter  fields in these models are exactly solvable. In particular they
give black-hole solutions that can be formed classically from collapsing
matter. To discuss the quantum aspects of black holes, one may at the
semi-classical level introduce a conformal anomaly term into the original
action that arises from quantizing the matter fields. Remarkably this
leads to Hawking radiation [\cghs]! Unfortunately all the models considered
so far exhibit indefinite Hawking radiation, or absence of a ground state,
which is believed to be a consequence of ignoring the back-reaction of
radiation on the metric. Moreover a full quantum discussion would require
quantizing all fields, including the metric field and the dilaton field.
Work in this direction have been discussed in [\cb,\alwis].

Although a direct quantization of the metric field in four-dimensional
space-time is difficult, in two-dimensional space-time it is possible to
successfully quantize the induced  gravity of Polyakov that arises from
matter conformal anomaly [\sltwo,\kpz]. Since, in the field theory
formulation of two-dimensional-induced gravity of Polyakov, quantum
regulator terms such as $R^2$ give non-trivial interactions in conformal
gauge, while in the light--cone gauge $R^2$ is a quadratic kinetic term,
it is arguably more advantageous to quantize the Polyakov gravity in the
light--cone gauge, as was done in the analysis of [\sltwo,\kpz]. For in
the CGHS-type of models the quantization of matter fields yields the
Polyakov-induced gravity, it is natural to think that it might also be
more advantageous to choose the  light--cone gauge instead of the
conformal gauge here.

In this paper we will carry out such an analysis in the light--cone gauge.
Section 1 contains the classical solutions of black holes,
found first in [\witten,\cghs], and their analogues in a
different coordinate system that corresponds to light--cone gauge choice
for the metric. In Sec 2, we will calculate, in the light--cone gauge, the
semi-classical matter stress tensor induced from the conformal anomaly (or
rather gravitational anomaly), which gives a rate of Hawking radiation that
agrees with that of the previous calculation. Finally we quantize the
metric field in the light--cone gauge {\it \`a la} KPZ [\kpz], but in
this case Fadeev-Popov ghosts that correspond to gauge-fixing couple to
a (flat)  metric, instead of the metric of the black-hole background
geometry. This  is necessary so as to avoid Hawking radiation of ghosts
[\stro]. We argue that this may serve as a useful framework to formulate
a full consistent quantum theory of dilaton gravity.



\noindent{\bf 2.\  Classical Black-Hole Solutions}

The action of dilaton gravity coupled conformal matter of the CGHS model
is  given by $S=S_D+S_M$ with
$$
\eqalign{
S_D &= {1\over 2\pi}\int d^2\sigma \sqrt{-g}
e^{-2\phi}\big[R+4(\nabla\phi)^2
+4\lambda^2\big],\cr
\noms
S_M &= -{1\over 4\pi}\int d^2\sigma \sqrt{-g}
\sum_{i=1}^N (\nabla f_i)^2.\cr}
\eqn\action
$$
Classically this theory is exactly solvable [\cghs]. It is conveniently
seen in conformal gauge, given by
$$
g_{++}=0=g_{--}, \qquad g_{+-}=g_{-+}=-{1\over 2} e^\rho,\eqn\cong
$$
with $\sigma^{\pm}=\sigma^0\pm\sigma^1$. In this gauge the action becomes
$$
S= {1\over \pi}\int
d^2\sigma\Big[e^{-2\phi}(-2\pa_+\pa_-\r +4\pa_+\phi\pa_-\phi -\lambda^2
e^{2\r}) -{1\over 2}\sum_{i=1}^N\pa_+f_i\pa_-f_i\Big]. \eqn\cact
$$
The equations of motion of dilaton and matter are given by
$$
\eqalign{
&-4\pa_+\pa_-\phi+4\pa_+\phi\pa_-\phi+2\pa_+\pa_-\r+\lambda^2 e^{2\r}=0,\cr
&\pa_+\pa_- f_i=0; \quad \qquad i=1,2,\cdots N.\cr}\eqn\cdf
$$
The $\r$ field equation of motion, corresponding to $T^{\rm c}_{+-}$,
\foot{The superscript c indicates that it is a classical stress tensor.} is
given by $$
T^{\rm c}_{+-}=e^{-2\phi}(2\pa_+\pa_-\phi
-4\pa_+\phi\pa_-\phi-\lambda^2 e^{2\r})=0.
\eqn\cpm
$$
In addition there are two constraints (that come from the equations of
motion of $g_{++}$ and $g_{--}$) given by
$$
\eqalign{
T^{\rm c}_{++} &=e^{-2\phi}(4\pa_+\r\pa_+\phi-2\pa_+^2\phi)
+{1\over 2}\sum_{i=1}^N\pa_+f_i\pa_+f_i=0,\cr
T^{\rm c}_{--} &=e^{-2\phi}(4\pa_-\r\pa_-\phi-2\pa_-^2\phi)
+{1\over 2}\sum_{i=1}^N\pa_-f_i\pa_-f_i=0.\cr}\eqn\cppmm
$$
Given an arbitrary solution $f_i=f_i^+(\sigma^+)+f_i^-(\sigma^-)$ of the
matter equations, the general solutions for $\r$ and $\phi$ are given by
$$
\crampest
\eqalign{
e^{-2\phi} =& {M\over\lambda}-\lambda^2\int d\sigma^+ e^{w_+}\int d\sigma^-
e^{w_-}-{1\over 2}\int d\sigma^+ e^{w_+}\int d\sigma^+ e^{-w_+}\sum_{i=1}^N
(\pa_+ f_i^+)^2\cr
&\  -{1\over 2}\int d\sigma^- e^{w_-}\int d\sigma^- e^{-w_-}\sum_{i=1}^N
(\pa_- f_i^-)^2, \cr
\noms
\r-\phi=& {w_+ +w_-\over 2},\cr}\eqn\sol
\uncramp
$$
where $w_+(\s^+)$ and $w_-(\s^-)$ are arbitrary gauge functions, and $M$ is
an integration constant.

Consider solutions with $f_i=0$. Fixing the residual conformal subgroup of
diffeomorphism in this gauge by setting $w_+=w_-=0$, one has the following
solution:
$$
e^{-2\phi}={M\over\lambda}-\lambda^2 \s^+\s^- =e^{-2\r},\eqn\bh
$$
up to constant shifts in $\s$. In the case when $M$ is non-vanishing, this
corresponds to a two-dimensional black hole with an ADM mass $M$
[\witten,\cghs], in which the event horizon is given by $\s^+\s^-=0$, and
the singularity lies behind it on $M/\l-\l^2\s^+\s^-=0$. The future time
infinity $i^+$, space infinity $i^0$ and past time infinity $i^-$
correspond to $(\infty,0)$, $(\infty,-\infty)$ and $(0,-\infty)$
respectively, while the past and future null infinity, ${\cal I}^-$
and ${\cal I}^+$, correspond to $\s^-\to -\infty$ and $\s^+\to\infty$
respectively.

When $M=0$, the metric goes  over to a flat metric
under the coordinate transformation $\s^{\pm}=\pm e^{\pm{\hat\s}^\pm}$,
while the dilaton field becomes linear in $\hat\s$. This is nothing but the
linear dilaton vacuum.

These black hole solutions with non-vanishing $M$ can be generated from the
linear dilaton vacuum as a result of matter perturbation [\cghs]. Take a
shock wave travelling in the $\s^-$ direction described by the stress tensor
$$
{\hat T}^f_{++}={1\over 2} \sum_{i=1}^N\pa_+f\pa_+f =a\delta(\s^+-\s^+_0).
\eqn\shock
$$
In the gauge $w_+=w_-=0$, one finds that
$$
e^{-2\phi}=-a(\s^+-\s^+_0)\Theta(\s^+-\s^+_0)-\l^2 \s^+\s^- =e^{-2\r},
\eqn\bhld
$$
where $\Theta(\s^+ -\s^+_0)$ is the step function.
This solution thus joins the linear dilaton vacuum together with a black
hole of mass $a\s^+_0\lambda$ along the line $\s^+=\s^+_0$ of the $f$ wave.

This classical analysis of black-hole solutions in the CGHS model can be
performed equally well in the light--cone gauge. Here we will carry it
out to set up the stage for a semi-classical analysis as well as a
full quantum treatment of the theory in the light-cone gauge, when the
metric field is quantized.

Consider the following gauge choice for the metric\foot{We use $x$ as
coordinate so as to dintinguish it from the $\s$ in the conformal gauge.}:
$$
ds^2=dx^+dx^- + h_{++} (dx^+)^2.\eqn\ligg
$$
 The action \action\  reduces to the following in this gauge choice:
$$
\eqalign{
S =&\ {1\over \pi} \int d^2 x \big[e^{-2\phi}(\pa_-^2 h_{++}+
4\pa_+\phi\pa_-\phi-4h_{++}(\pa_-\phi)^2+2\lambda^2)\big]\cr
&\ -{1\over \pi} \int d^2 x \Big[{1\over 2}\sum_{i=1}^N\pa_+f_i\pa_-f_i
+{1\over 2}h_{++}\sum_{i=1}^N(\pa_-f_i)^2\Big]. \cr}
\eqn\lact
$$
The dilaton and matter equations of motion are thus given by
$$
\crampest
\eqalign{
&e^{-2\phi}(\pa_-^2 h_{++}-4\pa_+\phi\pa_-\phi+4h_{++}(\pa_-\phi)^2
+4\pa_+\pa_-\phi-\lambda^2-4h_{++}\pa_-^2\phi-4\pa_-h_{++}\pa_-\phi)=0\cr
\noms
&\pa_-(\pa_+f_i-h_{++}\pa_-f_i)=0;\qquad i=1,2,\cdots N,\cr}\eqn\ldf
\uncramp
$$
while the $h_{++}$ equation of motion is given by
$$
T^c_{--}= -2e^{-2\phi}\pa_-^2\phi+{1\over 2}\sum_{i=1}^N(\pa_-f_i)^2=0,
\eqn\lmm
$$
supplemented by two constraints (eqs. of motion for $g_{+-}$ and $g_{--}$)
given by
$$
\crampest
\eqalign{
T^c_{+-}=&\  e^{-2\phi}(2\pa_+\pa_-\phi-2\pa_-h_{++}\pa_-\phi
-4h_{++}\pa_-^2\phi
-4\pa_+\phi\pa_-\phi+4h_{++}(\pa_-\phi)^2-\lambda^2)\cr
&+{1\over 2}h_{++}\sum_{i=1}^N(\pa_-f_i)^2 =0,\cr
\noms
T^c_{++}=&\  e^{-2\phi}(-2\pa_+^2\phi-2\pa_-h_{++}\pa_+\phi
+2\pa_-\phi\pa_+h_{++}
-4h_{++}\pa_-h_{++}\pa_-\phi\cr
\noss
&+8h_{++}\pa_+\pa_-\phi-8h_{++}^2\pa_-^2\phi-8h_{++}\pa_+\phi\pa_-\phi
+8h_{++}^2(\pa_-\phi)^2-2h_{++}\lambda^2)\cr
&-h_{++}\sum_{i=1}^N \pa_+f_i\pa_-f_i+h_{++}^2\sum_{i=1}^N(\pa_-f_i)^2
+{1\over 2}\sum_{i=1}^N \pa_+f_i\pa_+f_i=0.\cr}\eqn\lpmpp
\uncramp
$$
By certain linear combinations of \lpmpp, a set of simplified
equations reads as follows:
$$
\crampest
\eqalign{
&4e^{-2\phi}\pa_-^2\phi=\sum_{i=1}^N(\pa_-f_i)^2,\cr
\noms
&\pa_-(\pa_-h_{++}+2\pa_+\phi-2h_{++}\pa_-\phi)=0,\cr
\noms
&\pa_-^2 h_{++}+4\pa_+\phi\pa_-\phi-4h_{++}(\pa_-\phi)^2-\l^2=0,\cr
\noms
&\pa_+\pa_-f_i=\pa_-(h_{++}\pa_-f_i); \qquad  i=1,2,\cdots N,\cr
\noss
&\pa_+(e^{-2\phi}\pa_-^2 h_{++})
=h_{++}\pa_-\phi\sum_{i=1}^N\pa_+f_i\pa_-f_i
+{1\over 2}\pa_-\phi\sum_{i=1}^N(\pa_+f_i)^2
-h_{++}\pa_+\phi\sum_{i=1}^N(\pa_-f_i)^2.\cr}\eqn\sim
\uncramp
$$

The general solutions of \ldf--\lpmpp\  can presumably be given in
closed form. They are related to the general solutions in conformal gauge
\sol\  via an appropriate coordinate transformation. The gauge functions in
\sol\  will have their analogs in the general solutions in light--cone
gauge. The presence of these gauge functions reflects the fact that there
is a residual symmetry that leaves invariant the light--cone gauge choice
\ligg\  [\sltwo], given by
$$
\eqalign{
{\tilde x}^+ =&\ \alpha(x^+),\cr
{\tilde x}^- =&\ {x^-\over \alpha^\prime} +\beta(x^+),\cr}\eqn\res
$$
where $\alpha^\prime=d\alpha/dx$, while $h_{++}$ transforms as
$$
h_{++}=(\alpha^\prime)^2 {\tilde h}_{++} -{x^-\alpha^{\prime\prime}\over
\alpha^\prime}+\alpha^\prime\beta^\prime.\eqn\htrans
$$
The precise form of the general solutions is not of importance to us for
the moment. We will focus only on solutions of the type shown in  \bh\
in the absence of matter. In particular, solutions of our interest take the
following form:
$$
\eqalign{
\phi&=-{\l\over 2}(x^+ +x^-),\cr
h_{++}&= {M\over\l} e^{2\phi}. \cr}\eqn\lbh
$$
When $M$ is non-vanishing, this is exactly the black-hole solution of
mass $M$ in conformal gauge given in  \bh, which has been transformed
into the light--cone gauge via a coordinate transformation given by
$$
\eqalign{
x^+&={1\over\l}{\rm log}(\l \s^+),\cr
x^-&={1\over\l}{\rm log}({M\over\l}-\l^2 \s^+\s^-)-{1\over\l}{\rm log}(\l
\s^+).\cr} \eqn\trans
$$
Again the $M=0$ case corresponds to the linear dilaton vacuum.

Comparing the solutions \lbh\  with \bh, one finds that in light--cone
coordinates, the horizon lies on $\l(x^+ + x^-)={\rm log}(M/\l)$, and the
singularity on $x^+ + x^-=-\infty$, behind the horizon, where
the curvature $4\pa_-^2h_{++}$ indeed blows up. The future time infinity
$i^+$, space infinity $i^0$ and past time infinity $i^-$ are at
$(\infty,-\infty)$, $(\infty,\infty)$ and $(-\infty,\infty)$ respectively,
while the past and future null infinity, ${\cal I}^-$
and ${\cal I}^+$, correspond to $x^-\to\infty$ and $x^+\to\infty$
respectively.

\bigskip
\noindent{\bf 3.\  Semi-Classical Analysis}

It was proposed [\cghs] that quantum effects such as Hawking radiation
can be computed in a fixed-background geometry \bh\  by including the trace
anomaly term that arises from the quantization of the matter fields  $f$.
At the level of action it amounts to adding the well-known
Polyakov--Liouville term, which is often referred to as the induced action
for two-dimensional gravity.  It is non-local in terms of metric fields
given by
$$
S_{\rm ind}=-{c\over 96 \pi}\int\sqrt{-g} R{1\over \dal} R,\eqn\induce
$$
where $c$ represents the central charge of the quantized matter system and
the ghost system. As discussed in [\cghs], the inclusion of such a term at
the semi-classical level incorporates both Hawking radiation and its
back-reaction on the background geometry.

To discuss Hawking radiation, consider the solution of  \bhld, which is
generated by a shock-wave perturbation. It suffices to calculate the
expectation value of the energy-momentum tensor for the quantized matter
fields, denoted by $T^F$,\foot{Here $F$ indicates some quantum nature of
matter fields $f$, its physical meaning becomes clear presently.} in  an
asymptotic flat coordinate ${\bar\s}^\pm$  via, for instance,
$e^{\l{\hat\s}^+}=\l\s^+,e^{-\l{\hat\s}^-}=-\l\s^- -a/\l$.
In conformal gauge, where the induced action \induce\  goes over to a free
action $\ft{c}{12\pi}\int\pa_+\r\pa_-\r$, the well-known trace
anomaly relation gives:
$$
T^F_{+-}=-{c\over 12}\pa_+\pa_-\r.\eqn\trace
$$
$T^F_{++}$ and $T^F_{--}$ can be
obtained by the conservation law given by
$$
\pa_{\pm} T^F_{\mp\mp}+\pa_{\mp}
T^F_{+-} -2\pa_{\mp}\r\ T^F_{+-}=0.\eqn\cons
$$
They take the form given by
$$
T^F_{\pm\pm}=-{c\over 12}\big(\pa_{\pm}\r\pa_{\pm}\r-\pa_{\pm}^2\r+
t_\pm({\bar\s}^{\pm})\big), \eqn\ppmm
$$
where $t_\pm({\bar\s}^\pm)$ can be fixed by the boundary conditions.

Imposing that $T^F_{++}$ vanishes in the linear dilaton region, and there
is no incoming radiation, namely
$T^F_{++}\to 0, T^F_{+-}\to 0$ as ${\bar\s}^-\to -\infty$, one may
fix the form of $t_{\pm}({\bar\s}^\pm)$ to be given by
$$
t_+({\bar\s}^+)\to 0,\qquad t_-({\bar\s}^-)\to
-{1\over 4} c\l^2[1-(1+ae^{\l{\bar\s}^-}/\l)^2].\eqn\tpm
$$
Thus one reads off the values of the stress tensor for the outgoing
radiation as ${\bar\s}^+\to \infty$:
$$
\eqalign{
&T^F_{++}\to 0,\qquad T^F_{+-}\to 0\cr
\noms
&T^F_{--}\to {c\l^2\over 48}\Big[1-{1\over
(1+a e^{\l{\bar\s}^-}/\l)^2}\Big].\cr} \eqn\rad
$$
At future time infinity (${\bar\s}^-\to\infty$), $T^F_{--}$ approaches
the constant value $c\l^2/48$ [\cghs]. This gives the rate of Hawking
radiation.

This calculation can be performed equally well in light--cone gauge. The
trace anomaly relation and conservation law read:
$$
\eqalign{
\nabla_+ T^F_{--}&=-{c\over 24}\pa_-^3 h_{++},\cr
\noss
T^F_{+-}&=h_{++}T^F_{--}+{c \over 24}\pa_-^2 h_{++},\cr
\noss
T^F_{++}&=h_{++}T^F_{+-}+{c\over 12}h_{++}\pa_-^2h_{++}
-{c\over 48}(\pa_-h_{++})^2-{c\over 24}\pa_+\pa_-h_{++} + B(x^+) ,\cr}
\eqn\tracons
$$
where $\nabla_+\equiv \pa_+ -h_{++}\pa_- -2\pa_-h_{++}$, and $B(x^+)$ is an
arbitrary function.

For simplicity, let us consider the {\it fixed}-background geometry of
\bh.  One may calculate $T^F$ by solving \tracons.
With the parametrization of $h_{++}$
$$
h_{++}={\pa_+F\over \pa_-F}.\eqn\para
$$
$F$ is solved easily to be given by
$$
F(x^+,x^-)=F(\zeta); \qquad \zeta=-{M\over\l}e^{-\l x^+}+e^{\l x^-},\eqn\qf
$$
where $F$ is now an arbitrary function of $\zeta$. The general solution
of $T^F_{--}$ to \tracons\  can be written as a Schwartzian derivative of
$F$ with respect to $x^-$:
$$
T^F_{--}=-{c\over 24}\{F,x^-\},\eqn\fmm
$$
where the Schwartzian is given by
$$
\{F,x^-\}\equiv {\pa_-^3 F\over\pa_-F}
-{3\over 2}\Big({\pa_-^2 F\over\pa_-F}\Big)^2.\eqn\schwartz
$$
Using  \qf,  we obtain the following $T^F_{--}$:
$$
T^F_{--}={c\l^2\over 48}
\Big(1-2e^{2\l x^-}\{F,\zeta\}\Big);\eqn\tmm
$$
$T^F_{++}$ and $T^F_{+-}$ can also be solved in this background. For
completeness, they are  given by
$$
\crampest
\eqalign{
T^F_{+-}&= {Mc\l\over 16}e^{-\l x^+ -\l x^-}
-{Mc\l\over 24}e^{-\l x^+ +\l x^-}\{F,\zeta\},\cr
T^F_{++}&={M^2 c\over 8}e^{-2\l x^+ - 2\l x^-}
-{Mc\l\over 24}e^{-\l x^+ -\l x^-}-{M^2 c\over 24}
e^{-2\l x^+}\{F,\zeta\}+B(x^+).\cr}
\eqn\fpmpp
\uncramp
$$

The requirement that there is no incoming radiation can be satisfied by
choosing appropriate functions $F$ and $B(x^+)$ so that $T^F_{--}$,
$T^F_{+-}$ and $T^F_{++}$ vanish on the past null infinity ${\cal I}^-$.
On the future null infinity ${\cal I}^+$, the metric becomes flat,
$T^F_{+-}=0$, $T^F_{++}=0$, and $T^F_{--}$ is given by  \tmm\  with
$\zeta=e^{\l x^-}$. The rate of Hawking radiation is thus obtained by the
value of $T^F_{--}$ at future time infinity, {\it i.e.} $x^+\to \infty,
x^-\to -\infty$, in which case  $\zeta=0$. We thus have the desired value
$T^F_{--}=c\l^2/48$, provided that $\{F,\zeta\}$ is regular at $\zeta=0$.

The physical meaning of $F$ is that it is a semi-classical description of
the quantum matter fluctuation. This becomes clear once we compare the
parametrization \para,  which defines $F$ for certain $h_{++}$, with the
classical equation of motion for matter field $f$ in \ldf. However,
$F$ is not completely determined by Hawking radiation in a fixed background
geometry. This indeterminacy is reminiscent of the loss of information
and quantum coherence in the process of Hawking radiation [\hawk].

Next is the issue of the back-reaction of Hawking radiation on
the  metric. Semi-classically this can be discussed by solving a new
set of equations of motion that arise from the induced Polyakov--Liouville
action to the classical stress tensor. In the original CGHS model, the new
set of equations are not exactly solvable, and special kinds of solutions to
these equations, as well as numerical solutions, indicate [\rst-\hawkk,\hs]
that a singularity occurs. Models with additional conserved currents
[\cb--\rstt] do have solvable semi-classical equations of motion, but they
either suffer from a naked singularity or inevitably evolve into a
thunderbolt [\hs].

To address these issues in the light--cone gauge, one may start by writing
down the set of semi-classical equations. Here we will limit our
considerations to the case without classical matter $f$.
The dilaton equation of motion in \ldf\  is not
modified by the induced action. The matter equation in \ldf\  becomes null
in the absence of matter.
However, quantum matter fluctuation $F$ takes a form \para\  similar to the
classical equation of motion. The equation of motion
for $h_{++}$ \lmm\  and two constraints \lpmpp\  turn into
$$
T^{\rm c}+T^F=0,\eqn\eqsemi
$$
where the $T^{\rm c}$'s are those given \lmm, \lpmpp\  with $f$ turned off;
$T^F_{--}$ is given in  \fmm, while the $T^F_{++}$ and $T^F_{+-}$ are
related to the $T^F_{--}$ by \tracons, thus expressed in terms of $F$ as
follows:
$$
\crampest
\eqalign{
T^F_{+-}=&-{c\over 24}{\pa_+F(\pa_-^2F)^2\over(\pa_-F)^3}
-{c\over 12}{\pa_+F\pa_-^3F\over(\pa_-F)^2}
+{7c\over 24}{\pa_+\pa_-F\pa_-^2F\over(\pa_-F)^2}
+{c\over 24}{\pa_+\pa_-^2F\over\pa_-F},\cr
T^F_{++}=&{7c\over 24}{(\pa_+F)^2(\pa_-^2F)^2\over (\pa_-F)^4}
-{c\over 6}{(\pa_+F)^2\pa_-^3F\over (\pa_-F)^3}
-{c\over 8}{\pa_+\pa_-F\pa_-^2F\pa_+F\over (\pa_-F)^3}
+{c\over 6}{\pa_+F\pa_+\pa_-^2F\over (\pa_-F)^2}\cr
&\ \ +{c\over 24}{(\pa_+F)^2(\pa_-F)^2\over (\pa_-F)^2}
+{c\over 48}{(\pa_+\pa_-F)^2\over (\pa_-F)^2}
-{c\over 24}{\pa_+^2\pa_-F\over \pa_-F}.\cr}
\uncramp\eqn\fpmpp
$$
Exact solutions to \eqsemi\  would presumably be rather difficult to
find, and will be left for future work.

\bigskip
\noindent{\bf 4.\  Full Quantum Treatment}

To quantize the metric field $h_{++}$ in the light--cone gauge, we follow
 [\kpz]. The quantum analogues of the constraint equations \lpmpp\  are
rather tricky to implement. They can be done as follows [\kpz].
Changing from the gauge choice \ligg\  to a new gauge choice given by
$$
h_{--}=h_{--}(x), \qquad h_{+-}=h_{+-}(x), \eqn\gauge
$$
where $h_{--}$ and $h_{+-}$ are certain fixed but unspecified functions.
\foot{Note that they are not quantum fields like $h_{++}$.} Gauge
invariance   implies that the partition function ${\cal Z}$, as well as any
gauge     invariant quantity, must be independent of the choice of gauge
functions  $h_{--}$ and $h_{+-}$, and in  particular it implies that
$$
{\delta{\cal Z}\over \delta h_{--}}\Big|_{h_{--}=0}=0, \qquad
{\delta{\cal Z}\over \delta h_{+-}}\Big|_{h_{+-}=0}=0.\eqn\inv
$$
These conditions are just the quantum constraints that
$$
T^{\rm tot}_{++}=0, \qquad T^{\rm tot}_{+-}=0. \eqn\pppm
$$
The vanishing of $T^{\rm tot}_{++}$ requires that the total central charge
be zero.

In order to calculate the gravitational contribution $T_{++}[h_{++}]$ to
the total stress tensor, the authors of [\kpz] made use of an
observation by Polyakov [\sltwo] that there exists a hidden
$SL(2,R)$ Kac--Moody symmetry in the induced two-dimensional gravity.
The gravitational stress tensor $T_{++}[h_{++}]$ is constructed to be a
constrained Sugawara form in terms  of the currents of the underlying
$SL(2,R)$ Kac--Moody algebra. Thus the gravitational contribution the total
central charge is given by
$$
c(h_{++})=-6(k+2)-{6\over k+2} +15,\eqn\ch
$$
where $k$ is the level of the underlying $SL(2,R)$ Kac--Moody algebra.
The geometrical meaning of this $SL(2,R)$ symmetry is precisely the
residual symmetry of the light--cone gauge, given in \res\  and \htrans\
[\sltwo]. Since in the CGHS model in the light--cone gauge this residual
symmetry remains, we expect that there exists at least an $SL(2,R)$
symmetry. Thus we assume that \ch\  also holds in the present case.

It is necessary to introduce two pairs of Fadeev--Popov
ghosts $(\epsilon,\eta)$ and $(c,b)$, corresponding to the gauge-fixing
conditions in \gauge. The ghost contribution to  the total central
charge is a subtle issue, because of the fact that there is an
ambiguity in defining the metric to which the ghosts may couple. This
was addressed in  [\stro], and it was argued that it is
the (flat) metric $g^{\rm flat}=e^{-2\phi}g$  to
which the ghosts are coupled, so as to avoid Hawking radiation of ghosts.
Here we will proceed with this assumption\foot{Semi-classical
quantization of dilaton gravity considered in [\tanii] allows
more general ans\"atz.}. The gauge-fixing term is thus given by
$$
{\cal L}_{\rm gh}=b\nabla_- c +\eta(\nabla_+ c -\nabla_-\epsilon),
\eqn\ghos
$$
where $\nabla_\pm$ are covariant derivatives with respect to the flat
metric $g^{\rm flat}$.
Integrating out these ghost fields, one obtains the ghost
determinant that is essentially the induced action \induce\  for the flat
metric $g^{\rm flat}$, with an appropriate overall constant indicating that
the ghost central charge is $-26-2=-28$. Expressed in terms of $h_{++}$ and
$\phi$, it is thus given by $$ {7\over 12}\Gamma[h_{++}]
+{7\over 6\pi}\int (\phi \pa_-^2 h_{++}
+ \phi\pa_-(\pa_+-h_{++}\pa_-)\phi),\eqn\fadeev
$$
where $\Gamma[h_{++}]$ can be viewed as a gravitational analogue of WZW
action, given in a non-local form in $h_{++}$ by
$$
\Gamma[h_{++}]={1\over2\pi}\int\pa_-^2h_{++}{1\over\pa_-
(\pa_+-h_{++}\pa_-)}
\pa_-^2h_{++}.\eqn\gwzw
$$
Thus the second term of  \fadeev\  modifies
the kinetic term of the dilaton field $\phi$ and as well as introduces a
background charge for it.

We will denote the central charge that comes from the dilaton by $c_\phi$.
The matter central charge $c_{\rm matter}$ is $N$ in the present case with
$N$ scalars. The vanishing of the total central charge,
$0=c_{\rm tot}=c_{\phi}+N-28+c(h_{++})$, thus yields a relation, given by
\foot{The choice made here between the two roots of $k$  corresponds to
its correct classical limit as $N\to \infty$.}
$$
k+2={c_\phi-N-13+\sqrt{(c_\phi+N-1)(c_\phi+N-25)}\over 12}.\eqn\ktwo
$$
In the case of gravity without dilaton, the coefficient $k$ is nothing
but the overall renormalization constant for the quantum effective action
$\Gamma_{\rm eff}[h_{++}]$ [\kpz,\poly]. It is conjectured in  [\poly] that
the full quantum effective action to all loops for the induced gravity
\induce\  is given by
$$
\Gamma_{\rm eff}[h_{++}]={k\over 2}\Gamma\Big[{k+2\over k}h_{++}\Big],
\eqn\eff
$$
where $\Gamma[h_{++}]$ is given in \gwzw.
The renormalization constant $(k+2)/k$ of $h_{++}$ was checked by
performing a 1-loop calculation using either a determinant argument
[\poly]   or Feynman diagram perturbative expansion [\pk]\foot{Many of
these results of two-dimensional Polyakov
gravity have their analogues of higher-spin gravity, namely
$W$ gravity [\wgrav].}. We expect that these results
carry over to the present case of gravity with dilaton.

In [\cb,\alwis] it was shown in conformal gauge that a sequence of field
re-definition reduces the kinetic terms for the dilaton $\phi$ and the
conformal mode $\r$ to two free fields. One of which has a background
charge that dictates the string susceptibility of the theory [\tanii],
while the other contributes one unit to the central charge. In our
analysis, if we assume that $c_\phi=1$, then $k=(N-24)/6$ in the large $N$
limit.

A few remarks are in order. Owing to the correspondence between the KPZ
quantization in light--cone gauge [\kpz] and that of conformal field theory
adopted in  [\ddk], it is natural to think that there might exist a
conformal field theory for the full quantum dilaton gravity in conformal
gauge. Indeed this has been pursued in [\cb,\alwis], and further elucidated
in [\tanii]. It may well be that it is a necessary requirement to have a
conformal field description.

We have been rather cavalier in postulating the quantum implications \ch\
and \eff\  of the $SL(2,R)$ residual symmetry \res\  and \htrans. A rigorous
proof would require more careful analysis. In  [\sltwo], the
light-cone metric field $h_{++}$ there was expanded into
$h_{++}=J^+ -2J^0 x^- +J^- (x^-)^2$, in which $J^0,
J^\pm$ generate the underlying $SL(2,R)$ symmetry. In the present case, one
might hope that the metric field $h_{++}$ for the black-hole background
could be expanded similarly. This does not seem to be the case. A possible
choice is to expand the flat metric field $g^{\rm flat}$ that couples to
the ghosts, in which case the underlying $SL(2,R)$ generating current
describe quantum fluctuation around this flat background metric, instead of
the black-hole background.

A quantum treatment of the dilaton field is yet lacking so far. With the
assumption that the ghosts couple to the flat metric, quantization scheme of
dilaton will need to incorporate modification terms in \fadeev.  It is
plausible that dilaton quantization may also be dictated by some symmetry,
as that of the metric field by $SL(2,R)$. It may well be that there exist
some analogue of Polyakov's analysis in the presence of black-hole geometry
with the dilaton field. And one may discover bigger hidden symmetries, for
example, some $W$ symmetry advocated in  [\emn]. The exact physical meaning
and implication of these considerations are beyond the scope of the present
paper.

\bigskip
\noindent{\bf 5.\  Conclusions}

In this paper we have investigated the CGHS model in the light--cone gauge.
At  the classical level, solutions of black holes are found, corresponding
to   those found in conformal gauge via coordinate transformation. We have
calculated the rate of Hawking radiation, which agrees with the result in
conformal gauge. Other related issues, such as back-reaction, singularity
and information loss would be analysed in this gauge by looking for
solutions to \eqsemi\  and studying their behaviour. Quantization of the
metric $h_{++}$ has been discussed along the lines of [\kpz]. A proper
quantization of the dilaton field still needs to be incorporated into the
picture, before some of the fundamental issues of quantum gravity can be
better understood.

Other models [\cb--\rstt,\nooda] that are closely related to the CGHS one
can be studied similarly in the light--cone gauge. The models of
[\cb,\alwis] are presumably equivalent to the fully quantized CGHS
model, such as that quantized here in light--cone gauge. This wisdom
comes from the  equivalence between the KPZ [\kpz] and DDK [\ddk] formalism.
It would be very interesting to make this more transparent in the case of
two-dimensional dilaton gravity. The model of  [\rstt], with additional
conserved currents, may also be quantized in the light--cone gauge. It
would be very interesting to see the implications of the conserved currents,
if any, in this language.

\ack{I am very grateful to F.~ Bastianelli, E.~ Bergshoeff, A.~ Bilal,
P.~ Candelas, M.~J.~ Duff, L.~J.~ Romans, R.~ Williams and G.~ Zemba for
discussions and valuable help.}

\refout
\end